\documentclass[]{spie}  %>>> use for US letter paper
\usepackage[]{graphicx}

\title{A CNOT Gate in a Glass Chip} 

\author{Paul M. Alsing\supit{a}, Grigoriy Kreymerman\supit{b} and Warner A. Miller\supit{b}
\skiplinehalf
\supit{a} Air Force Research Laboratory, Information Directorate, Rome, NY 13441, USA; \\
\supit{b}Dept. of Physics, Florida Atlantic University, 777 Glades Road, Boca Raton, FL 33431, USA 
}

\authorinfo{Further author information: (Send correspondence to W.A.M.) E-mail: wam@fau.edu}
\begin{document} 
  \maketitle 

%%%%%%%%%%%%%%%%%%%%%%%%%%%%%%%%%%%%%%%%%%%%%%%%%%%%%%%%%%%%% 
\begin{abstract}
In our earlier work we posited that simple quantum gates and quantum algorithms can be designed utilizing the diffraction phenomena of a photon within a multiplexed holographic element. The quantum eigenstates we use are the photonÕs transverse linear momentum (LM) as measured by the number of waves of tilt across the aperture. Two properties of linear optical quantum computing (LOQC) within the circuit model make this approach attractive. First, any conditional measurement can be commuted in time with any unitary quantum gate; and second, photon entanglement can be encoded as a superposition state of a single photon in a higher--dimensional state space afforded by LM.  We describe here our experimental results for construction a controlled NOT (CNOT) gate logic within a holographic medium, and present the quantum state tomography for this device. Our theoretical and numerical results indicate that OptiGrateÕs photo-thermal refractive (PTR) glass is an enabling technology. This work has been grounded on coupled-mode theory and numerical simulations, all with parameters consistent with PTR glass. We discuss the strengths (high efficiencies, robustness to environment) and limitations (scalability, crosstalk) of this technology. While not scalable, the utility and robustness of such optical elements for broader quantum information processing applications can be substantial.  
\end{abstract}

%>>>> Include a list of keywords after the abstract 


%%%%%%%%%%%%%%%%%%%%%%%%%%%%%%%%%%%%%%%%%%%%%%%%%%%%%%%%%%%%%
\section{PHOTONIC QUANTUM ALGORITHMS WITHIN A VOLUME HOLOGRAM}
\label{sec:intro}  % \label{} allows reference to this section

In a previous manuscript we suggested that the construction of a small, lightweight, field deployable, inexpensive, low-dimensional, quantum computer is feasible using existing photonics technology \cite{Mil11c,Mil14}.  Here we demonstrate this approach experimentally by the construction of a CNOT gate. Photons offer great promise in quantum information processing (QIP) given their minimal decoherence.  However, it is this very resiliency that hinders their utility in quantum computing.  Their weak coupling with atomic structures leads to substantial inefficiencies. While they may not find themselves as the central facet  of a quantum CPU, we believe that they can, nevertheless, be an integrable part of the quantum information processing done by a quantum computer or in a quantum communication system where repetitive low-dimensional, fixed and robust  quantum algorithmic tasks are required (e.g. quantum error correction, quantum relays, quantum memory/CPU buss, ...).

Various approaches have been proposed for quantum computing.  Perhaps the most familiar of these is the quantum circuit model (QCM) \cite{NieChu00}.  Here the challenge is to identify a suitable subset of a universal set of quantum gates that can  faithfully represent a family of unitary operations on $d$ qubits ($d$-partite), such that the the number of gates required scale polynomially in $d$, whilst the dimension grows exponentially as $2^d$ .  After the unitary evolution the output needs to be projected onto the computational basis and irreversibly measured.  One realization of the QCM has been linear optical quantum computing (LOQC) and one-way quantum computing (OWQC), which are also known as cluster-state quantum computing \cite{KniLafMil01,RauBri01,Nie05,Kwi02}.  The essential feature of each of these approaches involves either a non-linear measuring process or a preparation of hyper-entangled input states ($|IN\rangle$), or both. Here a sequence of measurements are made to project the output state ($|OUT\rangle$)  onto one or another of the computational basis states, i.e onto a mutually unbiased basis (MUB).  In this manuscript we will explore the utility of phototnic quantum circuit models.

In this manuscript and in previous work\cite{Mil11c},  we suggest that in principle any quantum algorithm can be encoded within a single hologram, and that in practice, we demonstrate this here on a CNOT gate constructed by four independent recordings of the interference of superimposed transverse linear momentum (LM) photon states in a stack of four thin $1.64~mm$ holograms. In general, a qu{\it d}it state would require $d$ multiplexed recordngs of  the superposition of $2^d$-dimensional transverse LM states.  This thickness will ensure high fidelity unitary transformations \cite{Gle10}.  Nevertheless, there are commercially available holographic media, e.g. OptiGrate's photo-thermal refractive (PTR) glass,  that can support faithful holograms with a thickness approaching a few centimeters, thereby extending  our considerations to 10 to 20 dimensional quantum state spaces \cite{CiaGleSmi06}.  Every application of LOQC to quantum gates that we are aware of requires a cascade of interferometers which is highly unstable and impractical.  The approach here will ``lock"  these interferometers within a tempered piece of glass that is resistant to environmental factors.

Two properties of quantum computing within the circuit model make this approach attractive. 
First, any conditional measurement can be commuted in time with any unitary quantum gate - the timeless nature of quantum computing. 
Second, photon entanglement can be encoded as a superposition state of a single photon in a higher-dimensional state space.
While not scalable, the utility and robustness of such optical elements for broader quantum information processing applications can be substantial.

In this paper we describe an application of this LOQC holographic approach on the construction and tomographic measurement of such a multiplexed (stacked) volume hologram.    In Sec.~\ref{sec:vhg} we briefly outline the structure and function of a volume holographic grating that will be referred to  throughout this manuscript.   In Sec.~\ref{sec:lm} we describe the photonic  quantum eigenstates we use.  In particular we use the space spanned by the photon's linear momentum (LM) as measured by the number of waves of tilt across the aperture, i.e. the transverse component of the wave vector parallel to the face of the hologram.
 In Sec.~\ref{sec:cnot} we describe the construction of the familiar $CNOT$ gate without the need for multiplexing. Our alternative approach will ordinarily require stacking more holographic gratings. In Sec.~\ref{sec:exp} we describe our verification of such a CNOT gate. In conclusion, (Sec.~\ref{sec:sw}) we will  discuss the strengths (high efficiencies, robustness to environment) and limitations (scalability, crosstalk) of this technology. We emphasize here that the holographic approach presented here is not scalable, and with existing technology it is not quickly reprogrammable; nevertheless, the utility and robustness of such optical elements for broader quantum information processing applications can be substantial.  There should be many facets of a quantum computer other than the quantum CPU that could take advantage of this technology, including error correction, QKD relays, quantum memory management tasks and the like.

%%%%%%%%%%%%%%%%%%%%%%%%%%%%%%%%%%%%%%%%%%%%%%%%%%%%%%%%%%%%%
\section{VOLUME HOLOGRAPHIC GRATINGS} \label{sec:vhg}

Volume holography is used today for 2D image storage utilizing $394\, pixels/\mu m^2$, which consumes only $1\%$ of the theoretical volumetric storage density ($1/\lambda^3$) \cite{Bur01}.  This field was first introduced by Dennis Gabor in 1948 and it  has been a popular research field ever since the development of the laser in 1960.   Thicker holograms have more precise angular selectivity, i.e. its ability to differentiate the difference between two plane waves separated by a small angle,  and under certain well-known conditions  can achieve near perfect  efficiency \cite{Goo05}.  A  hologram is considered a volume hologram if its thickness $d \gg \Lambda^2/\lambda,$ where $\Lambda$ is the characteristic period of the index of refraction of the grating, and $\lambda$ is the wavelength of the light.  It is important for our purposes to emphasize that volume holography enables higher storage densities, and under suitable recording configurations can achieve near perfect efficiencies. This is not possible with thin holograms and  planar gratings (spatial light modulators) where efficiencies can be at most $33\,\%$.

\begin{figure}
   \begin{center}
   \begin{tabular}{c}
   \includegraphics[height=6cm]{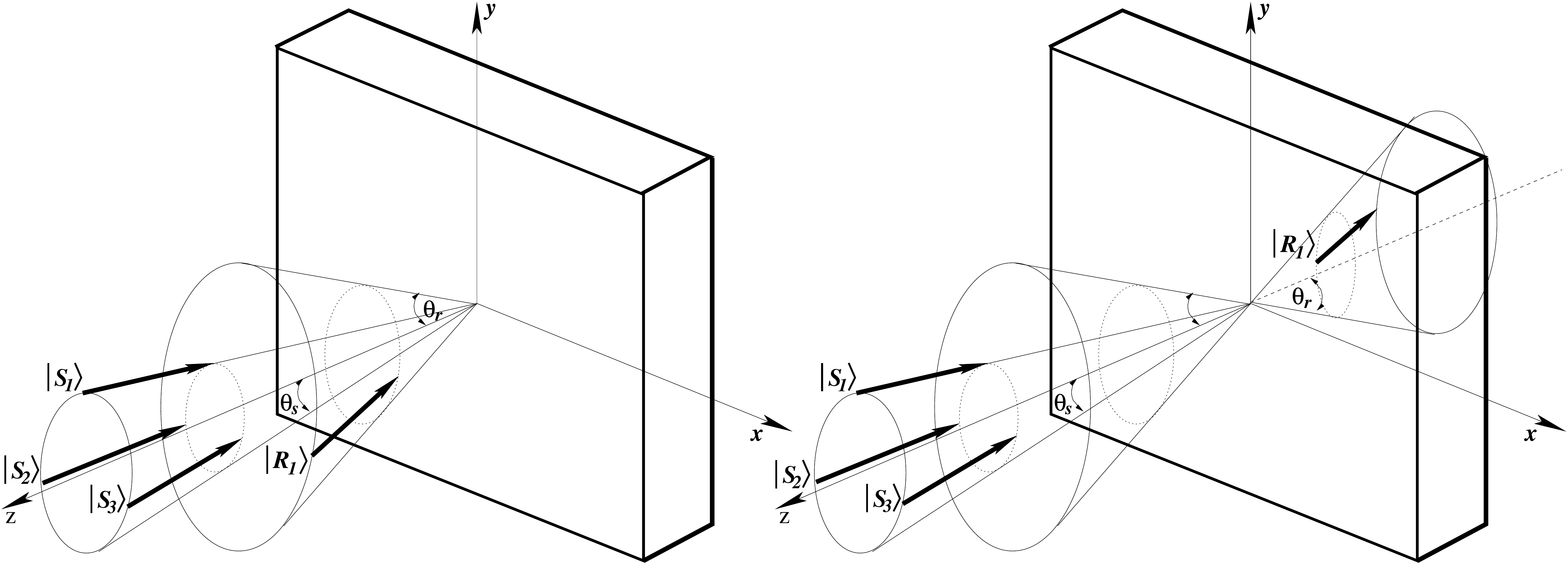}
   \end{tabular}
   \end{center}
   \caption[example]
%>>>> use \label inside caption to get Fig. number with \ref{}
   { \label{fig:vhg}
   The left diagram shows a  recording of a volume transmission grating by the coherent superposition of a plane reference wave
   $|R_1\rangle$ and a linear superposition of three signal waves
   $|S\rangle = e^{i \alpha_1} |S_1\rangle + e^{i \alpha_2} |\beta S_2\rangle + e^{i \alpha_3} | S_3 \rangle$ .
   On the right we show the function of the hologram. If the identically oriented signal wave $|S\rangle$ is sent
   into the hologram then the reference wave, $| R_1\rangle$, will be reconstructed in the diffraction.
   %The diffraction pattern will ordinarily consist of the other four modes corresponding to the components of the signal wave;
   The diffraction pattern will ordinarily consist of higher order diffracted modes parallel to the signal state.
   However, for a suitably tuned volume hologram perfect efficiency can be achieved, as shown in the right diagram \cite{Mil11}.  This is why we  constrained the signal wave components to a cone of half angle $\theta_s$ centered on the normal to the hologram face. }
   \end{figure}

Briefly, the transmission volume holograms we consider here are  formed when a ``signal" wave, $\langle \vec r | S \rangle= A(\vec r) e^{i\Phi(\vec r)}$ is directed into a holographic material and made to coherently interfere with an oblique  "reference" plane wave, $|R\rangle$ as illustrated in the left diagram of  Fig.~\ref{fig:vhg} for  $N=3$.  In the figure the ``signal" wave is a superposition of $N$ plane waves,
\begin{equation}
\langle \vec r |S\rangle = \sum_{i=1}^N \langle \vec r | S_i \rangle =   \sum_{i=1}^N e^{i\alpha_i} e^{i \vec k_i \cdot \vec r},
\end{equation}
where $\alpha_i$ are pure phase angles.  In this paper,  we only consider planar reference waves,  and the signal state as the superposition of plane waves.  Ordinarily the signal waves will have variable phase and amplitude modulations.  After the hologram is developed, and if we direct the identical signal wave, $\langle \vec r | S\rangle$,  into the hologram, then for a perfectly tuned hologram,  the reference plane wave, $\langle \vec r | R\rangle$,  should emerge from  the hologram as illustrated in the right diagram of Fig.~\ref{fig:vhg}.  If it is not tuned, then other diffracted orders, e.g.  modes parallel to the signal states, may emerge.

How does one form a perfectly tuned hologram within the coupled mode theory? We have shown recently \cite{Mil11} that near perfect efficiencies can be obtained if (1) the hologram thickness is tuned to its optimal thickness, (2) if the each of the signal's Fourier wave vectors have the same projection onto the normal to the hologram surface, i.e. they all lie on a cone with half angle $\theta_s$ as shown in Fig.~\ref{fig:cone}, and (3) each of the reference plane waves lie on their own distinct cone concentric with the first, with half angle $\theta_r$ and centered on the normal to the hologram face.   We will also consider multiplex holograms wherein multiple independent exposures are made within the holographic material before it is developed.  We  demonstrated  using coupled-mode theory that if the  signal waves $\{S_i\}_{i,1,2, ...N}$  form an orthogonal set under the $L_2$ norm in the plane perpendicular to the waves propagation direction ($z$), i.e.
\begin{equation}
\langle S_i | S_j \rangle = \int S^*_i(x,y) S_j(x,y) dxdy = \delta_{ij},
\end{equation}
then  perfect efficiency can be achieved for each of the signals \cite{Mil11,Goo05}.
%-------------
%-------------

A volume multiplexed hologram that has achieved perfect efficiency (within coupled-mode theory \cite{Kog69}) under the ``3+1"  conditions outlined above provides a linear map between signal and reference modes. Physically it represents a projection (or redirection) operator or signal state sorter,\cite{Mil11}
\begin{equation} \label{eq:po}
\hat{\cal  P} = | R_1 \rangle\langle S_1 | + | R_2 \rangle\langle S_2 | + \ldots + | R_N \rangle\langle S_N|,
\end{equation}
uniquely identifying each pair of signal and reference waves.

Although the index of refraction within the hologram can be rather complicated, these devices are strictly linear optical components.  Therefore, the diffraction patterns for a beam of photons will correspond exactly to the probability distribution for a single photon in the beam.  For the remainder of this manuscript we will assume we are dealing with low number Fock states.  We will describe in the next two sections (1) how we encode a $d$-partite quantum state onto a single photon that can be used in volume holography, and (2) how the unitary operation representing an entire quantum algorithm can be encoded in a multiplexed hologram.

%%%%%%%%%%%%%%%%%%%%%%%%%%%%%%%%%%%%%%%%%%%%%%%%%%%%%%%%%%%%%
\section{ENCODING A $D$-PARTITE STATE ON A SINGLE PHOTON} \label{sec:lm}

The dimension, $N$, of a quantum state space spanned by the direct product of $d$ qubits grows exponentially, $N=2^d$.  Thus to represent a single $d$-partite state by a single photon requires vastly more quantum numbers than available to its two polarization degrees of freedom.  However, the photon can be characterized by its extrinsic properties as characterized by its wavefront  amplitude and phase. The potential of extending photon-based quantum information processing and quantum computing to higher dimensions was made possible in 1983 when Miller and Wheeler \cite{MilWhe84} described a fundamental  set of quantum experiments utilizing a photon's orbital angular momentum (OAM), and when Allen et~al \cite{All92} reintroduced Laguerre-Gaussian light beams that possessed a quantized orbital angular momentum (OAM) of $l\hbar$ per photon.   This opened up an arbitrarily high dimensional quantum space to a single photon \cite{MilWhe84,AllBarPad03}.  Following these discoveries Mair et~al. \cite{MaiVazWeiZel01,Oem04} unequivocally demonstrated the quantum nature of photon OAM by showing that pairs of OAM photons can be entangled using the non-linear optical process of parametric down conversion. Shortly thereafter, Molina-Terriza et~al. \cite{MolTorTor02} introduced a scheme to prepare photons in multidimensional vector states of OAM commencing higher-dimensional QIP, with applications to quantum key distribution. Recently a practical method has been demonstrated to produce such MUB states using computer-generated holography with a single spatial light modulator (SLM) \cite{Gru08}.

While photons with specific values of OAM have been emphasized recently in the literature,  one can equally well utilize any other extrinsic set of orthogonal basis functions for higher-dimensional QIP, e.g. energy, linear momentum, angular momentum.  While OAM states respect azimuthal symmetry, transverse or linear momentum (LM) states respect rectilinear symmetry.  Independent of the representation we use, the mutually unbiased (MUB) states will ordinarily be modulated in both amplitude and phase\cite{Gru08}.

While the advantage of higher-dimensional QIP lies in its ability to increase bandwidth while simultaneously tolerating a higher bit error
rate (BER) \cite{Cer02},  such transverse photon wave functions are more fragile to decoherence under propagation than the photon's spin wave function \cite{Pat05}, and the divergence of the states in propagation may require larger apertures.  However, since the spatial distances within a quantum computer are small this is not a concern here.  This is not the case for quantum key distribution.  Our preliminary results show that the efficiencies one can achieve with holographic elements exposed by LM plane waves and their superpositions (volume Bragg gratings) is substantially higher than that for photon wavefronts exhibiting OAM \cite{Mil11,Boy,Gun}.  Multiplexed volume Bragg gratings have been thoroughly addressed in the literature \cite{Kog69,Cas75,CiaGleSmi06,MohGayMag80}.  Since the efficiencies are potentially higher, and volume Bragg gratings are well understood, we therefore will restrict our attention in this paper to the orthonormal transverse photon states of LM.

\begin{figure}
   \begin{center}
   \begin{tabular}{c}
   \includegraphics[height=7cm]{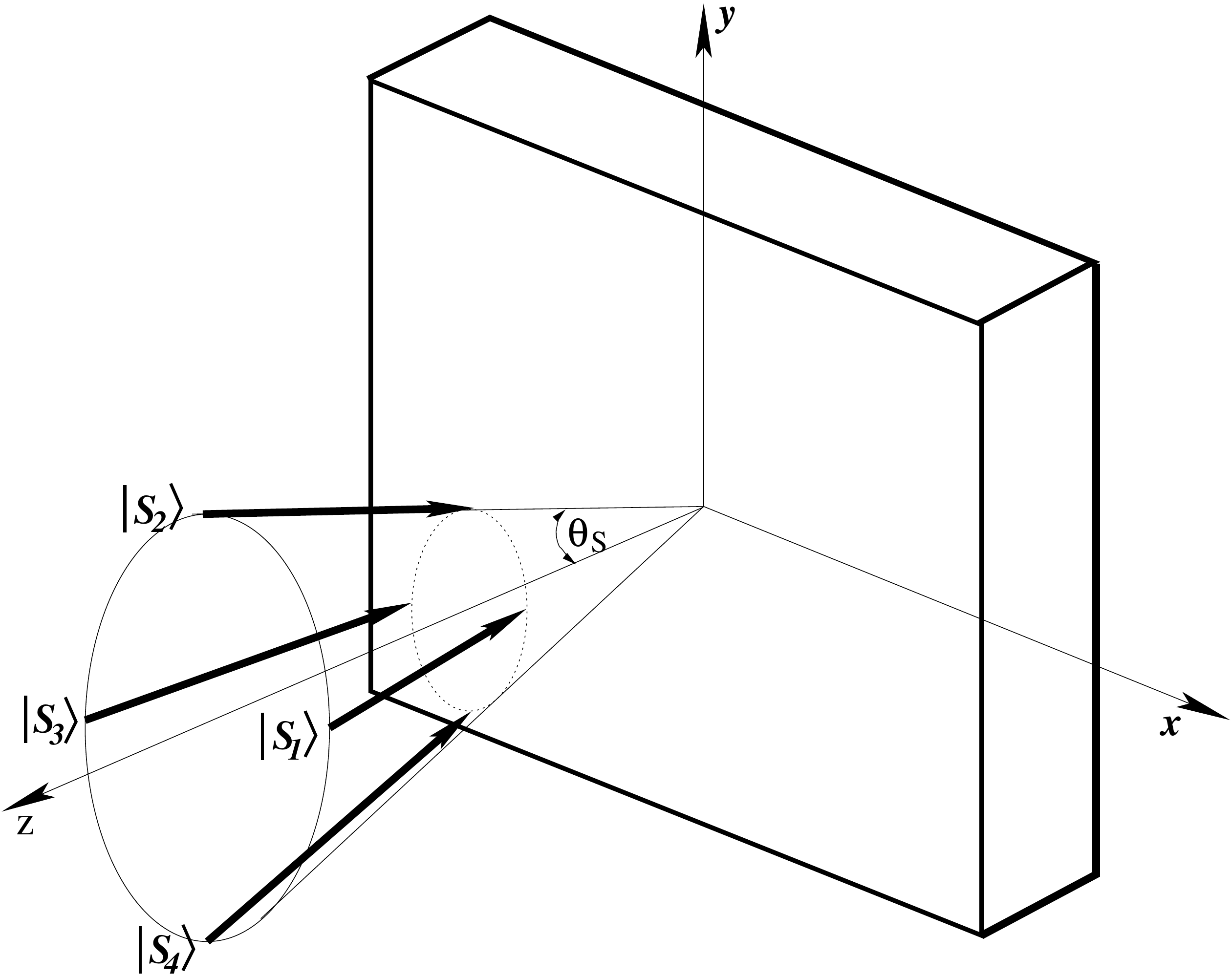}
   \end{tabular}
   \end{center}
   \caption[example]
%>>>> use \label inside caption to get Fig. number with \ref{}
   { \label{fig:cone}
  These Four LM states (arrows) that lie on a cone are used to generate a bipartite, or  4-dimensional signal state space.   Here we chose the eigenstates (each a plane wave) in this 8-dimensional space with eigenvectors, $|S_i\rangle = |\theta_s,\phi_i\rangle,\ i\in{1,2,3,4}$, here $\phi_1=0$, $\phi_2=\pi/2$, $\phi_3=\pi$ and $\phi_4=3\pi/2$.  This can be generalized to higher dimensions, $2^d$, by introducing $d$ eigenstates around the cone, as long as the angular selectivity of the hologram is less than $\Delta \phi=2\pi/2^d$. The wave vectors are defined in Eq.~\ref{eq:momentum}. }
   \end{figure}

We  concentrate in this section on the physics of transverse or LM states.  We quantize the LM within the  2-dimensional plane of the face of the hologram,  Fig.~\ref{fig:cone}.  We know that Bragg gratings can obtain perfect efficiencies as long as the component perpendicular to the face of the hologram  ($k_\perp$) for each of the photons' momentum are the same i.e. the plane wave states must lie on a cone.  In addition the thickness of the hologram must be carefully chosen \cite{Kog69,Goo05} .

%-------------
%-------------

If we are to represent a $d$-partite state by a single LM photon, we will require at least $N=2^d$ LM states around the cone of half angle $\theta_s$ centered on the normal to the hologram's face.  As a result, we can freely choose as our  computational quantum state basis any $2^d$ non-colinear plane waves. In this case, our integer quantum numbers will be the azimuthal angles, $\phi_i,\ i\in\{1,2,\ldots 2^d\}$ locating the wave vector of the photon around the cone.   Here we assume the hologram surface is in the $x$-$y$ plane.  For photons of wavenumber $k$ these waves correspond to a transverse linear momentum $p_\parallel =  \sin{(\theta_s)} \hbar k = \hbar k_\parallel$, with components $p^x_i\ =  \sin{(\phi_i)} \hbar k_\parallel$,    $p^y_i\ =  \cos{(\phi_i)} \hbar k_\parallel$ and $p^z_i  = \cos{(\theta)} \hbar k = \hbar k_\perp = constant$; respectively.  Here,  $k_\parallel = k \left( \lambda/D\right) $ is the magnitude of the transverse component of the wave vector of a plane wave of wavelength $\lambda$  with one "wave of tilt"  ($\lambda/D$) across the aperture of breadth $D$.

In the frame of the hologram,  and in units where the speed of light is
unity, the components of the 3-momentum, $\vec p = \{p^x, p^y, p^z\}$,  for each of our $2^d$ photons can be expressed in terms of their wavenumber, ($k$), i.e.,
\begin{equation}\label{eq:momentum}
\vec p_{i} =  \hbar \vec k_i =   \hbar \ \left\{k^x_i,k^y_i,k^z_i\right\} =
\hbar k \ \left\{\sin{(\theta)} \cos{(\phi_i)},\sin{(\theta)} \sin{(\phi_i)}, \cos{(\theta_s)}\right\}.
\end{equation}
These $N=2^d$ plane wave states, $\langle \vec r| S_i\rangle = e^{i \vec k_i \cdot \vec r}$   define our {\em computational basis} for  quantum information processing.
\begin{equation}
\label{eq:basis}
MUB = \left\{ | S_1 \rangle, | S_2 \rangle, \ldots,  | S_{N} \rangle \right\}.
\end{equation}
Each of these states represents a transverse Fourier mode of a photon;
they are orthogonal under the $L^2$ norm ($\langle S_i | S_j \rangle = \delta_{ij}$) and
span our $2^d$-dimensional state space.

%%%%%%%%%%%%%%%%%%%%%%%%%%%%%%%%%%%%%%%%%%%%%%%%%%%%
\section{THE CNOT GATE: STACKING INSTEAD OF MULTIPLEXING} \label{sec:cnot}

\begin{figure}
   \begin{center}
   \begin{tabular}{c}
   \includegraphics[height=4in]{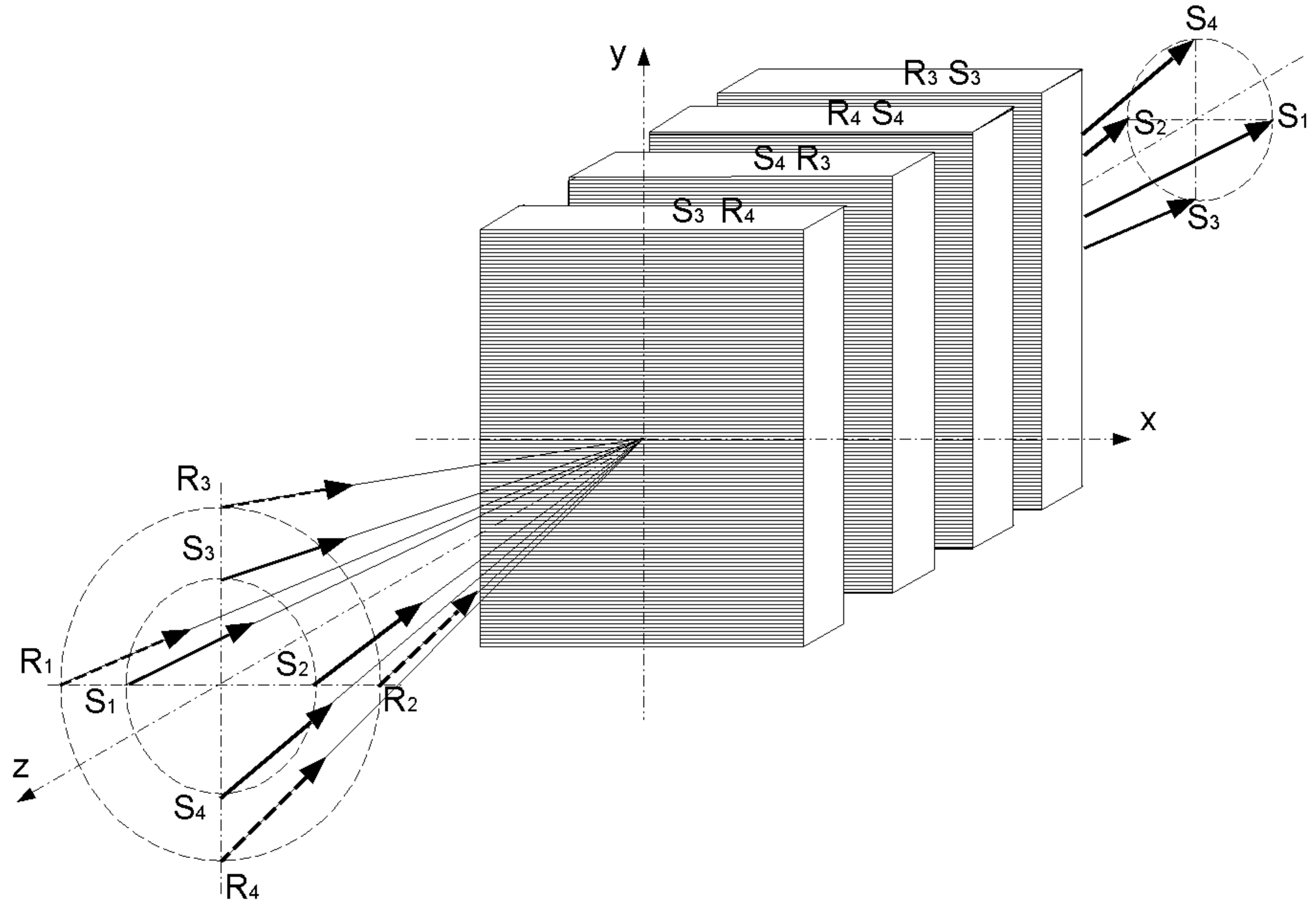}
   \end{tabular}
   \end{center}
   \caption[example]
%>>>> use \label inside caption to get Fig. number with \ref{}
   { \label{fig:cnot}
Volume holographic design of the 4-dimensional CNOT gate in PTR glass.  The gate can be constructed by a stack of 4 LM gratings, or by a stack of two multiplexed gratings. }
   \end{figure}

One alternative to multiplexing is to make single recordings in each of many holograms and to stack the holograms and is essentially equivalent.  Here we provide an illustrative example with a specific design of a quantum $CNOT$ gate compatible with  PTR glass.  This gate is realized by stacking four holograms, and we will describe each below.  The $CNOT$ gate is a two qubit gate. Therefore the dimension of the state space is  4-dimensional. While this state space can be constructed as a product space of qubits by utilizing the polarization states of two correlated photons, it can also be represented by a single LM photon in a 4-dimensional state space.  The  $CNOT$ gate can be constructed with a single photon.  Following the arguments of the previous section, we freely choose four independent plane waves lying on the cone shown in Fig.~\ref{fig:cnot}.  We associate to these independent transverse LM modes the four orthogonal quantum state vectors, $|S_1\rangle$, $|S_2\rangle$, $|S_3\rangle$ and $|S_4\rangle$.  Any state vector, $|\psi\rangle$, in this  4-dimensional state space can be written as a linear superposition of these normalized states,
\begin{equation}
|\psi\rangle = \alpha |S_1\rangle + \beta |S_2\rangle + \gamma |S_3\rangle + \delta |S_4\rangle,
\end{equation}
with,
\begin{equation}
|\alpha|^2 + |\beta|^2+|\gamma|^2 + |\delta|^2 = 1.
\end{equation}
Each of our basis states can be expressed in matrix notation,
\begin{equation} \label{eq:8}
|S_1\rangle = \left( \begin{array}{c} 1 \\ 0 \\ 0\\ 0 \end{array} \right), \
|S_2\rangle = \left( \begin{array}{c} 0 \\ 1 \\ 0\\ 0 \end{array} \right), \
|S_3\rangle = \left( \begin{array}{c} 0 \\ 0 \\ 1\\ 0 \end{array} \right), \  \hbox{and}\
|S_4\rangle = \left( \begin{array}{c} 0 \\ 0 \\ 0\\ 1 \end{array} \right).
\end{equation}
In this computational basis the CNOT gate can be expressed by the following unitary transformation:
\begin{equation} \label{eq:ucnot}
CNOT = \left( \begin{array}{c c c c}
1 & 0 & 0 & 0 \\
0 & 1 & 0 & 0 \\
0 & 0 & 0 & 1 \\
0 & 0 & 1 & 0
 \end{array} \right).
\end{equation}
If we let the $z$-axis be orthogonal to the face ($x$-$y$ plane) of the hologram, the four volume holographic gratings are recorded by a suitable superposition of the set of four signal plane waves,
\begin{equation}
\langle  \vec r | S_1\rangle = \exp\left(i \vec k_1 \cdot \vec r \right), \ \langle  \vec r | b\rangle = \exp\left(i k_2 \cdot \vec r \right), \
\langle  \vec r | S_2\rangle = \exp\left(i \vec k_3 \cdot \vec r \right), \  and \ \langle  \vec r | d\rangle = \exp\left(i k_4 \cdot \vec r \right),
\end{equation}
and four reference waves,
\begin{equation}
\langle  \vec r | R_1\rangle = \exp\left(i \vec \kappa_1 \cdot \vec r \right), \ \langle  \vec r | R_2\rangle = \exp\left(i \vec \kappa_2 \cdot \vec r \right), \
\langle  \vec r | R_3\rangle = \exp\left(i \vec \kappa_3 \cdot \vec r \right), \  and \ \langle  \vec r | R_4\rangle = \exp\left(i \vec \kappa_4 \cdot r \right),
\end{equation}
as shown in Fig.~\ref{fig:cnot}.

%-------------
%-------------

The hologram is recorded so that each row of the unitary matrix of the $CNOT$ gate  is used to generate its  own volume holographic grating.  For a 2-qubit gate such as the $CNOT$ gate we would ordinarily require four recordings; however, since the first two bits are just an identity matrix we need only two layers to transform the signal states into the desired reference states. In addition to one holographic recording per dimension of the state space,  we also require the conjugate of the each grating (two in the case of the CNOT gate) in order  to transform the diffracted reference waves from the reference waves back into the desired signal states. In particular, the CNOT-gate constructed from four holographic gratings stacked together as is shown in Fig.~\ref{fig:cnot}.
\begin{enumerate}
\item The first grating is recorded with the two coherent plane waves corresponding to states $|S_3\rangle$ and  $|R_4\rangle$ .
\item The second grating is recorded with the two coherent plane waves corresponding to states $|S_4\rangle$ and  $|R_3\rangle$ .
\item The third grating is recorded with the two coherent plane waves corresponding to states $|R_4\rangle$ and  $|S_4\rangle$ .
\item The fourth grating is recorded with the two coherent plane waves corresponding to states $|R_3\rangle$ and  $|S_3\rangle$ .
\end{enumerate}
The four gates will not diffract the first two signal states $|S_1\rangle$ or $|S_2\rangle$. However, the first two gratings redirect the two signal states $|S_3\rangle$ and $|S_4\rangle$ into $|R_4\rangle$ and $|R_3\rangle$ respectively, in  accordance with the Pauli $X$-gate,
\begin{equation}
\hat X = \left(  \begin{array}{c c} 0 & 1 \\ 1 & 0 \end{array}\right).
\end{equation}
The first hologram is equivalent to the operator,
\begin{equation}
\hat{\cal U}_{1} = |S_1 \rangle\langle S_1| + |S_2\rangle\langle S_2|+|R_4\rangle\langle S_3| + |S_4\rangle\langle S_4|,
\end{equation}
and the second hologram recorded with signal plane wave, $\langle \vec r | S_3 \rangle$ and reference plane wave,
$\langle \vec r | R_4 \rangle$, is equivalent to the operator,
\begin{equation}
\label{eq:cnotu2}
\hat{\cal U}_{2} = |S_1 \rangle\langle S_1| + |S_2\rangle\langle S_2|+|R_4\rangle\langle R_4| + |R_3\rangle\langle S_4|.
\end{equation}
After the signal state , $|IN\rangle$, passes through the first hologram the basis vectors change from $\{S_1, S_2, S_3, S_4\}$ to the orthonormal  basis, $\{S_1, S_2, R_4, S_4\}$. This explains why the third term in Eq.~\ref{eq:cnotu2} is entirely within the reference space.  

While these two recordings could have been made in a single multiplexed hologram, we recover the same function by stacking the two together, thereby generating the $CNOT$ operation,
\begin{equation}
\hat{\cal U}_{CNOT} = \hat{\cal U}_{2} \hat{\cal U}_{1} =  |S_1 \rangle\langle S_1| + |S_2\rangle\langle S_2|+|R_3\rangle\langle S_4| + |R_4\rangle\langle S_3|.
\end{equation}
However, the output of these two stacked holograms are the reference states, $|R_1\rangle$,  $|R_2\rangle$,  $|R_3\rangle$ and  $|R_4\rangle$. In order to redirect these back to the proper signal states, we require the redirection operator. This can be accomplished by recording a third hologram with the states $|R_3\rangle$ and $|S_3\rangle$. The third hologram is equivalent to the operator,
\begin{equation}
\hat{\cal U}_{3} = |S_1 \rangle\langle S_1| + |S_2\rangle\langle S_2|+|S_3\rangle\langle R_3| + |R_4\rangle\langle R_4|.
\end{equation}
Similarily, the fourth hologram is recorded with the states $|R_4\rangle$ and $|S_4\rangle$ and the fourth hologram is equivalent to operator,
\begin{equation}
\hat{\cal U}_{4} = |S_1 \rangle\langle S_1| + |S_2\rangle\langle S_2|+|S_3\rangle\langle S_3| + |S_4\rangle\langle R_4|.
\end{equation}
Therefore, the combination of the four stacked volume holograms have the desired action -- the CNOT gate.
\begin{equation}
\widehat{CNOT} = \hat{\cal U}_2 \cdot \hat{\cal U}_1\cdot\hat{\cal U}_2 \cdot \hat{\cal U}_1 = |S_1 \rangle\langle S_1| + |S_2\rangle\langle S_2|+|S_4 \rangle\langle S_3| + |S_3\rangle\langle S_4|.
\end{equation}
One can apply these principles to design a universal set of quantum gates, as well as simple quantum algorithms such as QT.

The advantage of stacking the holograms is that one can make the hologram thicker, thereby increasing the efficiency; however, achieving and maintaining the proper alignment should be only slightly more problematic. By multiplexing, we would need two holograms, each with two independent recordings in them.  The first would be equivalent to the last two holograms in Fig.~\ref{fig:cnot}, while the second would be equivalent to the first two and would just redirect the reference beams into their corresponding signal states. The first two recordings are complementary to the second two -- thus in some sense we are recording the ``square root" of the CNOT gate.

%%%%%%%%%%%%%%%%%%%%%%%%%%%%%%%%%%%%%%%%%%%%%%%%%%%%%%%%%%%%%
\section{Experimental Results: the Tomography of a Volume Holographic LM CNOT Gate} 
\label{sec:exp}

\begin{figure}
   \begin{center}
   \begin{tabular}{c}
   \includegraphics[height=5cm]{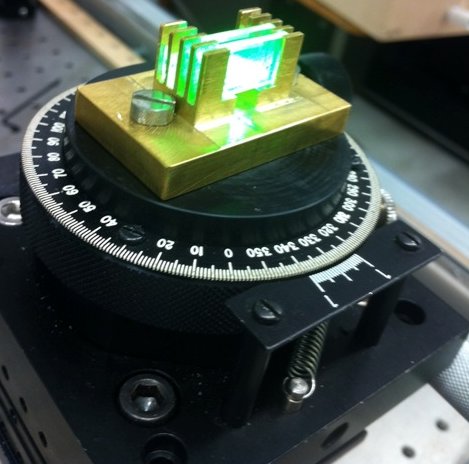}
   \end{tabular}
   \end{center}
   \caption[example] 
%>>>> use \label inside caption to get Fig. number with \ref{}
   { \label{fig:photo} 
Photograph of the stacked volume holograms in PTR glass illuminated by a Coherent Inc. 532nm green diode laser.}
   \end{figure}

Here we describe the experimental realization of the CNOT gate described above by measuring the quantum state tomography of this stacked-holographic gate as shown in Fig.~\ref{fig:photo}. Since this is a linear optical element, the tomographic map could be made using a non-attenuated continuous wave (CW) laser beam.  There was not need to verify this gate at the single photon/low Fock state level. Recall, the 2 qu{\it b}it entanglement is modeled by an equivalent linear superposition in a 4-dimensional quantum space of transverse LM of Eq.~\ref{eq:8}. 
%%  Use following command to specify that graphics file is in 
%%  a directory other than this LaTeX source file.
%%  Note use of / to separate subdirectories, for UNIX and Windows OS.
%%\graphicspath{{H:/HANSON/SPIESTY/}}
%% tabular environment useful for creating an array of images  
%-------------

\begin{figure}
   \begin{center}
   \begin{tabular}{c}
   \includegraphics[height=5cm]{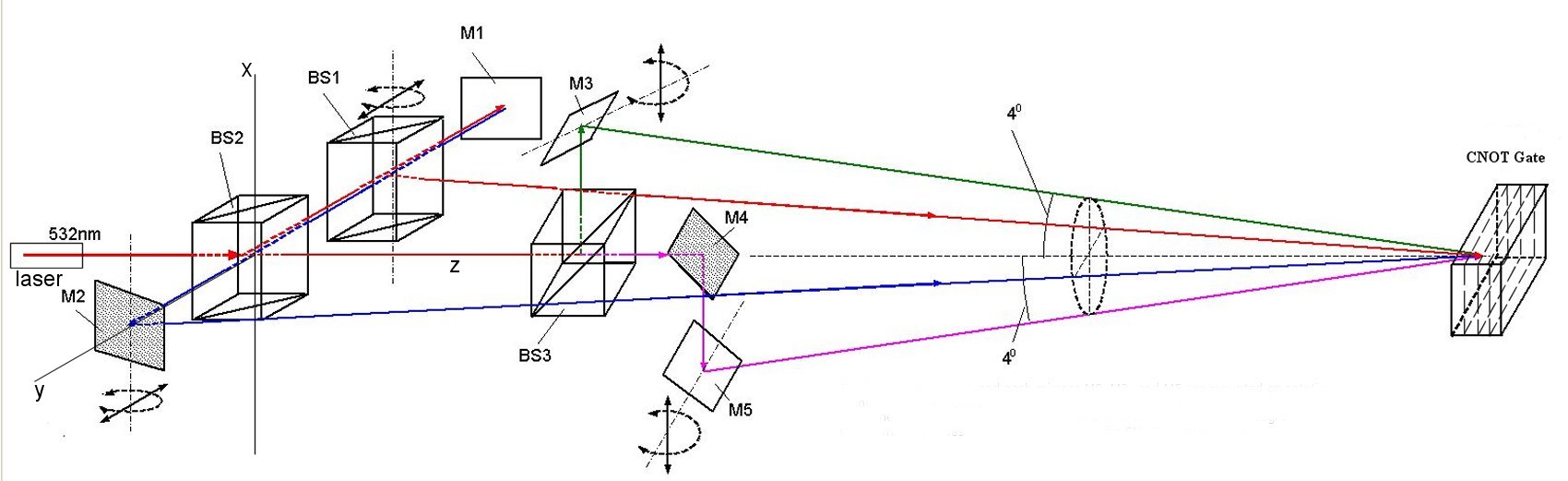}
   \end{tabular}
   \end{center}
   \caption[example] 
%>>>> use \label inside caption to get Fig. number with \ref{}
   { \label{fig:table} 
We illustrate here one of the possible optical arrangements we used to explore the tomography of the holographic CNOT gate. Here we employed three beam splitters (shown as BS) and five mirrors labeled as (M) in order to construct the required superposition of the four transverse LM states. We used a Coherent Inc. green diode laser.}
   \end{figure}

\begin{figure}
   \begin{center}
   \begin{tabular}{c}
   \includegraphics[height=7cm]{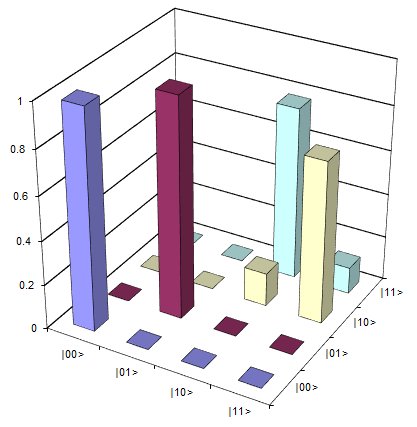}
   \end{tabular}
   \end{center}
   \caption[example] 
%>>>> use \label inside caption to get Fig. number with \ref{}
   { \label{fig:tom} 
The tomography of our CNOT gate. The bars in graph represent relative diffraction efficiency. This is the ratio between intensity in diffraction maximum and total intensity of light passing diffraction gratings. We measured the total intensity by placing a photometer in close proximity to grating to capture the majority of the light.  Subsequently, we positioned the photometer  to capture only light from diffraction maximum. We also repeated this for the states $|00\rangle$ and $|01\rangle$ where we predicted no diffraction. This was verified by our measurements. The observed intensity was 99\% and the remainder of the light, $\sim$1\%, was due to scattering (imperfections gratings material). The  remaining two basis states, $|10\rangle$ and $|11\rangle$, were not unitary as expected. The sum over the last two columns should ideally be unity; however, the sums are  0.93 and 0.85; respectively. These numbers are smaller since the gratings were not constructed with high optimization for the diffraction efficiency.  Our measurements fell within our estimate given the specifications of the volume Bragg gratings. }
   \end{figure}

The experimental results by way of the quantum state tomography is obtained from a CNOT gate fabricated from four stacked holographic gratings are shown in Fig.~\ref{fig:tom} and in Table~\ref{table:1}. These graphically reflect the unitary matrix shown in Eq.~\ref{eq:ucnot}  for the CNOT gate and physically reflects the  relative intensity of the diffracted light passing through the fabricated CNOT gate.  In particular, we used a C.W. Solid-State laser (COMPAS 315M-150) with wavelength 532nm and intensity 150 Mw used in our experiment. We used a  stack of four holographic grating in order to simulate CNOT gate as shown in Fig.~\ref{fig:photo}. The thickness of each grating is 1.64~{\em mm}.  The first two gratings are identical. Their Bragg angles at the wavelength 532nm are 4.41¡ and -23.61¡ (as measured relative to z-axis in z-y plane).  Their  diffraction efficiency (DE) are 88\%, acceptance angle at full with at half maximum (FWHM) 2.4mrad (Bragg selectivity) at 4.41¡ and a 2.6~mrad (Bragg selectivity) at -23.61¡. The sign minus indicates that the angle is on the opposite side of z-axis.  The last two volume holograms are also identical volume Bragg gratings.  Their characteristics are as follows: Bragg angles at a wavelength 532nm are 3.51¡ and 23.54¡; diffraction efficiency (DE) 93\%; acceptance angle at FWHM 2.9~mrad (Bragg selectivity) at 3.51¡ and 3.3~mrad (Bragg selectivity) at 23.54¡.  Each of the two sets of two gratings were  prepared from single holographic grating by cutting one thicker grating in half. The beams {\em S1} in Fig.~\ref{fig:table} reflected from mirror M3 is related to state $|00\rangle$, the beam {\em S2} reflected from mirror {\em M5} is related to state $|01\rangle$, the beam {\em S3} reflected from mirror {\em M2} is related to state $|10\rangle$, and finally, the beam {\em S4} reflected from {\em BS1} is related to state $|11\rangle$. The letters on the top gratings (Fig.~\ref{fig:table}) label the various beams that were used for recording the holographic Bragg gratings. There was  crosstalk of 15\% and 12\% between states $|10\rangle$ and $|11\rangle$ in our holographic gate. We expected this since in the manufacturing process the Bragg angles for two sets of gratings not precisely matched. In addition, the diffraction efficiency of the gratings was not optimized to the maximum possible value (about 99\%) and the gratings did not have any antireflection coating.  The goal of our experiment was to demonstrate a predictable application of this principle for one non-trivial popular gate.  We believe we demonstrated this.  We could have worked hard to obtain thicker holograms with more refined Bragg selectivity, tuned to give near perfect efficiency; but this was not our objective here. We wanted to observe, understand, model and controll the Bragg mismatch, lower efficiency and cross talk. 
            
In our experiment the beam splitter {\em BS1} and mirrors {\em M2} were mounted on rotation stage in $y$-$z$ plane and a one-dimensional vertical translation stage along $y$ axis.  The mirrors {\em M3}, and {\em M5} were mounted on a rotation stage in $x$-$z$ plane with a one-dimensional translation stage along $x$ axis. The beam splitters {\em BS2}, {\em BS3} and mirrors {\em M4} were installed so as to redirect the beam of light at a right angle, the mirror {\em M1} reflected the  light. The four beams which were reflected from {\em BS1}, {\em M2}, {em M3} and {\em M5}, can be adjusted to the relevant Bragg angles. The all these beams are directed into the hologram on a cone (Fig.~\ref{fig:cone} with semi angle, $\theta_s \approx 4\deg$. 

%%  Use following command to specify that graphics file is in 
%%  a directory other than this LaTeX source file.
%%  Note use of / to separate subdirectories, for UNIX and Windows OS.
%%\graphicspath{{H:/HANSON/SPIESTY/}}
%% tabular environment useful for creating an array of images  
%-------------
%-------------
%% This table is carefully placed in the source file to make 
%% it appear at bottom of page, but above the footnotes.  
%% Use of [h] in following command forces table to appear "here".
\begin{table}[h]
\caption{The tomography of the 4-stacked volume Bragg CNOT gate.} 
\label{table:1}
\begin{center}       
\begin{tabular}{|c|c|c|c|c|} %% this creates two columns
%% |l|l| to left justify each column entry
%% |c|c| to center each column entry
%% use of \rule[]{}{} below opens up each row
\hline
{\bf Input/Output} & $|00\rangle$ & $|01\rangle$ & $|10\rangle$ & $|11\rangle$   \\
\hline
 $|00\rangle$  & 0.99 & 0 & 0 & 0 \\
\hline
$|01\rangle$  & 0 & 0.99 & 0 & 0 \\
\hline
$|00\rangle$  & 0 & 0 & 0.15 & 0.73 \\
\hline
$|00\rangle$  & 0 & 0 & 0.78 & 0.12 \\
\hline
\end{tabular}
\end{center}
\end{table} 

These results are comparable to other attempts to produce a CNOT gate using conventional optical bench elements. For example, the results above are comparable to those observed by Pittman et al. \cite{Pit03}. The single mode solid state laser with wavelength 532 nm and 100 mW power was use in experiment. CNOT gate constructed from four experimental holographic Bragg gratings fabricated in  Photo-induced Processing Laboratory at UCF. The experimental setup, as one possible arrangement, for the test fabricated CNOT gate shown on Fig.~\ref{fig:table}. 
%%  Use following command to specify that graphics file is in 
%%  a directory other than this LaTeX source file.
%%  Note use of / to separate subdirectories, for UNIX and Windows OS.
%%\graphicspath{{H:/HANSON/SPIESTY/}}
%% tabular environment useful for creating an array of images  
%-------------
   
%%%%%%%%%%%%%%%%%%%%%%%%%%%%%%%%%%%%%%%%%%%%%%%%%%%%
\section{DISCUSSION: ADVANTAGES AND LIMITATIONS} \label{sec:sw}

Constructing simple quantum algorithms and quantum gates in volume holograms provides substantially greater optical stability  than the equivalent multi cascaded interferometric optical bench realizations.  For LOQC, this stability is the overarching advantage.  Often quantum operators, e.g. the simple projection operator given by  Eq.~\ref{eq:po}, require a cascade of interferometers where the output of one interferometer is used as  the input of the next \cite{Kwi02,Lea04, Zou05}. Therefore, as the dimension of each state space increases,  it becomes exceedingly hard to stabilize and is simply impractical beyond two qu{\it b}its.  Other approaches, such as crossed thin gratings lack the efficiency needed for QIP. The CNOT device constructed and benchmarked  achieved this in a single piece of glass without the problem of misalignment.  The technology presented here can potentially replace "fixed" optical components for a broad spectrum of classical and quantum photonics experiments.

The primary limitation of volume holographic QIP is that it is not scalable.  Experience shows that multiplexing requires approximately $1 mm$ per recording of the state space to achieve high fidelity, and in QIP applications this scales exponentially with the number of qubits. But then again, we are not aware of a realistic scalable quantum computer to date. Secondly,  the holograms discussed here are write-once holograms and cannot be erased. Therefore the algorithm is ``fixed" into the holographic element.  While there are re-recordable holographic media, none that we know of has the specifications to outperform PTR glass for the applications discussed in this manuscript.   This is hardly a prescription for a quantum CPU; however, as mentioned in the paper, this technology might be integral to complete QIP systems where smaller $d$-partite operations are needed on a routine basis, e.g. a quantum memory bus, quantum error correction circuit, QKD relay system, etc...

We continue to explore three questions: (1) how many independent writes of orthogonal states into a holographic element can be made in the PTR glass before cross-talk or saturation between the modes becomes a limiting factor? (2) Is it difficult to stack the holograms due to the enhanced angular selectivity of the volume holograms? And (3) what is the maximum number of recordings in a multiplexed PTR hologram that can be reasonably be achieved? Our results here on the simple CNOT gate give us encouragement for future applications. 

%%%%%%%%%%%%%%%%%%%%%%%%%%%%%%%%%%%%%%%%%%%%%%%%%%%%%%%%%%%%%
\acknowledgments     %>>>> equivalent to \section*{ACKNOWLEDGMENTS}       
 
We acknowledge the support and guidance of Leonid Glebov and in supplying the volume holographic elements. WAM acknowledges support from AFRL/RITC under grant FA 8750-10-2-0017, and from the ARFL's Visiting Faculty Research Programs.  Any opinions, findings and conclusions or recommendations expressed in this material are those of the author(s) and do not necessarily reflect the views of AFRL. 

%%%%%%%%%%%%%%%%%%%%%%%%%%%%%%%%%%%%%%%%%%%%%%%%%%%%%%%%%%%%%
%%%%% References %%%%%

%\bibliography{report}   %>>>> bibliography data in report.bib
%\bibliographystyle{spiebib}   %>>>> makes bibtex use spiebib.bst

%%%%%%%%%%%%%%%%%%%%%%%%%%%%%%%%%%%%%%%%%%%%%%%%%%%%%%%%%%%%%
%%%%% References %%%%%

%\bibliography{report}   %>>>> bibliography data in report.bib
%\bibliographystyle{spiebib}   %>>>> makes bibtex use spiebib.bst
%PMA comment: spiebib is not giving me refernces 23-27, so stop using article.bbl
%             and just include the bibliography directly!

%\section*{References}

\end{document}